\documentclass{aastex61}

\submitjournal{PASP}

\shorttitle{The Automatic Observation Management System: I}
\shortauthors{Han et al.}

\begin{document}

\title{An Automatic Observation Management System of the GWAC Network I: System Architecture and Workflow}

\correspondingauthor{Yujie Xiao}
\email{yjxiao@nao.cas.cn}
\correspondingauthor{PinPin Zhang}
\email{ppzhang@nao.cas.cn}

\author[0000-0002-6107-0147]{Xuhui Han} 
\affiliation{Key Laboratory of Space Astronomy and Technology, National Astronomical Observatories, Chinese Academy of Sciences, Beijing 100101, China.}

\author{Yujie Xiao}  
\affiliation{Key Laboratory of Space Astronomy and Technology, National Astronomical Observatories, Chinese Academy of Sciences, Beijing 100101, China.}

\author{PinPin Zhang}  
\affiliation{Key Laboratory of Space Astronomy and Technology, National Astronomical Observatories, Chinese Academy of Sciences, Beijing 100101, China.}

\author{Damien Turpin}  
\affiliation{Universit\'e Paris-Saclay and Universit\'e de Paris, CNRS, CEA, AIM, F-91191 Gif-sur-Yvette, France.}

\author{Liping Xin} 
\affiliation{Key Laboratory of Space Astronomy and Technology, National Astronomical Observatories, Chinese Academy of Sciences, Beijing 100101, China.}

\author{Chao Wu} 
\affiliation{Key Laboratory of Space Astronomy and Technology, National Astronomical Observatories, Chinese Academy of Sciences, Beijing 100101, China.}

\author{Hongbo Cai} 
\affiliation{Key Laboratory of Space Astronomy and Technology, National Astronomical Observatories, Chinese Academy of Sciences, Beijing 100101, China.}

\author{Wenlong Dong} 
\affiliation{Key Laboratory of Space Astronomy and Technology, National Astronomical Observatories, Chinese Academy of Sciences, Beijing 100101, China.}

\author{Lei Huang} 
\affiliation{Key Laboratory of Space Astronomy and Technology, National Astronomical Observatories, Chinese Academy of Sciences, Beijing 100101, China.}

\author{Zhe Kang} 
\affiliation{Changchun Observatory, National Astronomical Observatories, Chinese Academy of Sciences, Changchun 130117, China.}

\author{Nicolas Leroy} 
\affiliation{LAL, Univ Paris-Sud, CNRS/IN2P3, Orsay, France.}

\author{Huali Li} 
\affiliation{Key Laboratory of Space Astronomy and Technology, National Astronomical Observatories, Chinese Academy of Sciences, Beijing 100101, China.}

\author{Zhenwei Li} 
\affiliation{Changchun Observatory, National Astronomical Observatories, Chinese Academy of Sciences, Changchun 130117, China.}

\author{Xiaomeng Lu} 
\affiliation{Key Laboratory of Space Astronomy and Technology, National Astronomical Observatories, Chinese Academy of Sciences, Beijing 100101, China.}

\author{Yulei Qiu} 
\affiliation{Key Laboratory of Space Astronomy and Technology, National Astronomical Observatories, Chinese Academy of Sciences, Beijing 100101, China.}

\author{Jing Wang} 
\affiliation{Guangxi Key Laboratory for Relativistic Astrophysics, School of Physical Science and Technology, Guangxi University, Nanning 530004, China.}
\affiliation{Key Laboratory of Space Astronomy and Technology, National Astronomical Observatories, Chinese Academy of Sciences, Beijing 100101, China.}

\author{Xianggao Wang} 
\affiliation{Guangxi Key Laboratory for Relativistic Astrophysics, School of Physical Science and Technology, Guangxi University, Nanning 530004, China.}

\author{Yang Xu} 
\affiliation{Key Laboratory of Space Astronomy and Technology, National Astronomical Observatories, Chinese Academy of Sciences, Beijing 100101, China.}

\author{Yuangui Yang} 
\affiliation{School of Physics and Electronic Information, Huaibei Normal University, Huaibei 235000, China.}

\author{Yong Zhao} 
\affiliation{Key Laboratory of Optical Astronomy, National Astronomical Observatories, Chinese Academy of Sciences, Beijing 100101, China.}

\author{Ruosong Zhang} 
\affiliation{Key Laboratory of Space Astronomy and Technology, National Astronomical Observatories, Chinese Academy of Sciences, Beijing 100101, China.}

\author{Weikang Zheng} 
\affiliation{Department of Astronomy, University of California, Berkeley, CA 94720-3411, USA.}

\author{Yatong Zheng} 
\affiliation{Key Laboratory of Space Astronomy and Technology, National Astronomical Observatories, Chinese Academy of Sciences, Beijing 100101, China.}

\author{Jianyan Wei} 
\affiliation{Key Laboratory of Space Astronomy and Technology, National Astronomical Observatories, Chinese Academy of Sciences, Beijing 100101, China.}



\begin{abstract}
The Ground Wide Angle Camera Network (GWAC-N) is an observation network composed of multi-aperture and multi-field of view robotic optical telescopes. The main instruments are the Ground Wide Angle Cameras Array (GWAC-A). Besides, 
several robotic optical telescopes with narrower field of views (than the GWAC-A) provide fast follow-up multi-band capabilities to the GWAC-N. The primary scientific goal of the GWAC-N is to search for the optical counterparts of Gamma Ray Bursts that will be detected by the Space Variable Object Monitor (SVOM) satellite. The GWAC-N performs many other observing tasks including the follow-ups of Target of Opportunities (ToO) and both the detection and the monitoring of variable/periodic objects as well as optical transients. To handle all of those scientific cases, we designed 10 observation modes and 175 observation strategies, especially, a joint observation strategy with multiple telescopes of the GWAC-N for the follow-up of gravitational wave (GW) events.  To perform these observations, we thus develop an Automatic Observation Management (AOM) system in charge of the object management, the dynamic scheduling of the observation plan and its automatic broadcasting to the network management and finally the image management. The AOM system combines the individual telescopes into a network and smoothly organizes all the associated operations. The system completely meets the requirements of the GWAC-N on all its science objectives. With its good portability, the AOM is scientifically and technically qualified for other general purposed telescope networks. As the GWAC-N extends and evolves, the AOM will greatly enhance the discovery potential for the GWAC-N. In the first paper of a series of publications, we present the scientific goals of the GWAC-N as well as the hardware, the software and the strategy setup to achieve the scientific objectives. The structure, the technical design, the implementation and performances of the AOM system will be also described in details. In the end, we summarize the current status of the GWAC-N and prospect for the development plan in the near future.

\end{abstract}
\keywords{automated telescopes, observation management system (AOM)}




\section{Introduction}\label{s:intro}

In the last decade, a new type of network emerged. Thanks to the modern computing and communication technologies, these telescopes are designed to form a general-purpose observation network, such as Las Cumbres Observatory Global Telescope (LCOGT, Brown et al. 2013), the Global Relay of Observatories Watching Transients Happen (GROWTH, Kasliwal et al. 2019), the All-Sky Automated Survey for Supernovae (ASAS-SN, Shappee et al. 2014),  the Robotic Optical Transient Search Experiment (ROTSE, Akerlof et al. 2003),
the Pan-STARRS Survey (Chambers et al. 2016), the Rapid Action Telescope for Transient Objects (TAROT, Bo$\ddot{e}$r et al. 1999), and the Master-Net (Lipunov et al. 2010)... Such inclusions of individual facilities into a global interconnected network is a key to largely enhance the discovery potentials and take up the challenge of the multi-messenger astronomy of the next decade.  However, netting telescopes, organizing and scheduling for the general-purpose network are the common problems in modern observational astronomy, since these facilities with limited resource are designed only for given purposes, which requires different size, photometry parameters and controlling technics. A huge human intervention is still involved in the schedule process for most modern observation networks (Mora \& Solar 2010).

Under the framework of the Chinese-French Space Variable Object Monitor (SVOM) mission, an array consisted of a set of 9 Ground-based Wide-Angle Cameras (GWAC-A, hereafter) is designed to simultaneously search for the optical prompt emission of Gamma Ray Bursts (GRBs) detected by the SVOM on-board gamma-ray instruments (ECLAIRs and GRM,  Cordier et al. 2015; Wei et al. 2016). Furthermore, several robotic, multi-band, small Field of View (FoV) telescopes are also deployed for automatically validating and following up candidates detected by the GWAC-A. In fact, combining these wide FoV telescopes and fast-slewing, multi-band, small FoV telescopes in a well organized network can permits to obtain a better observational coverage and detection performances useful for multiple tasks such as, large-sample surveys, periodic and quasi-periodic objects, transient targets, moving objects. But successfully performing these observations depends not only on instrument properties but also on a network combining robotic telescopes, communication, observation scheduling, observation controlling and data processing. Besides, in order to catch the nature of different scientific targets, different optimized observation strategies must be implemented. The optimization of observation strategy should balance between the fruitful scientific returns and the limited telescope resource.  Therefore, we develop an automatic observation management system to integrate the facilities and software into a network named as the GWAC network (GWAC-N hereafter), since the GWAC-A, is the majority of this network. The automatic observation management system contains functions of observation target management, fully-automated dynamic observation scheduling and autonomous telescope dispatching, data management.  The system enhances the efficiency of uses of the GWAC-N to a great level by automatized carrying out multi-target, multi-telescope, simultaneous joint observations, all the routine observations for each telescope, and keeping the manual observing function. With standard datalink, this system can be easily adapted to other similar subjects in time-domain astronomy and extended to the collaborative telescopes.

\begin{figure}
\centering
\includegraphics[width=0.95\columnwidth]{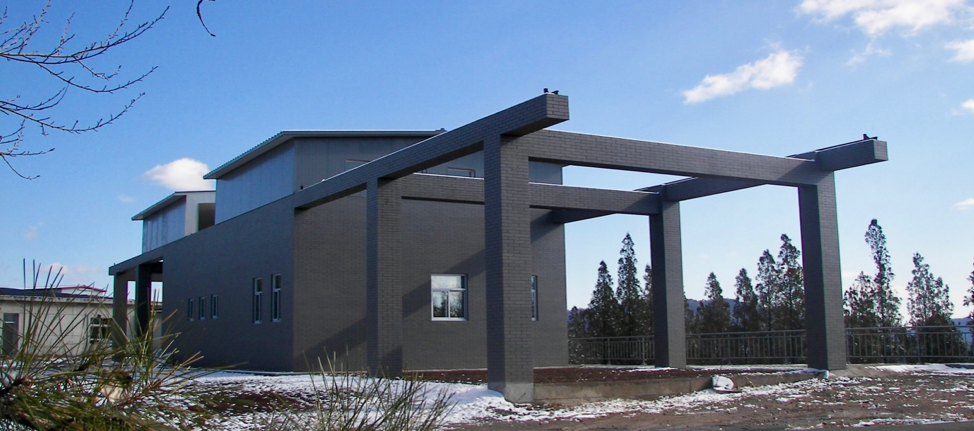}\\
\includegraphics[width=0.95\columnwidth]{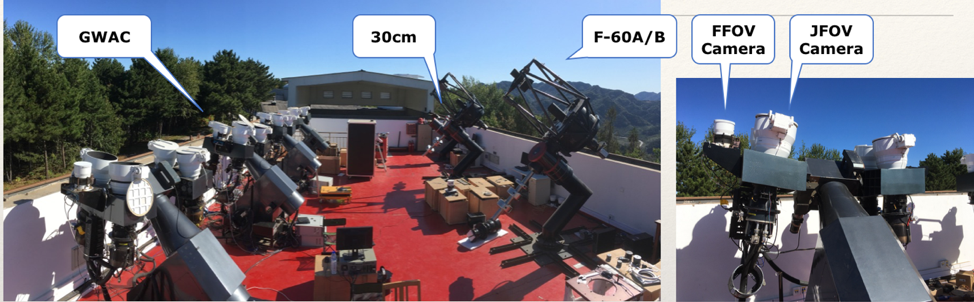}
\caption{Top: The dome of the GWAC. The roof can be opened along the rails. 
Bottom left: the telescopes of the GWAC-A, the GWAC-F60A/B and the GWAC-F30.
Bottom right: two types of cameras (JFoV and FFoV camera) mounted on one GWAC mount.
  }
\label{Fig:gwac_n_telescopes}
\end{figure}

In 2014, 12 mini-GWAC, the pathfinder telescopes of the GWAC-A started operation in the GWAC dome at Xinglong observatory (Huang et al. 2015). Two 60-cm follow-up telescopes (GWAC-F60A/B) were installed in 2015 and achieved first light in the same year. The first GWAC mount equipped with 4 Joint Field of View (JFoV) cameras and 1 Full Field of View (FFoV) camera were installed and tested in 2017. In 2018, 2 fully equipped GWAC mounts, 2 GWAC-F60A/B, 1 GWAC-F30 were in operation. Figure \ref{Fig:gwac_n_telescopes} shows the dome and telescopes of the GWAC-N. Although the telescopes were in place, they were still not connected as a network. The telescopes were operated separately and manually controlled by two observation assistants during night observations. Responding speed and observation efficiency were low. Observation capability for scientific targets was limited.  Paving a small part of sky with the GWAC-A and monitoring several targets with the GWAC-F60A/B and the GWAC-F30 were the pattens of the routine observations at this stage. Thus, the automatic observation management (AOM) system had been developed in 2019 to integrate the hardware and software of the GWAC-N to fulfill the scientific requirements described in the Section \ref{s:scientific_goal}. In the late 2019 and early 2020, the Tsinghua-NAOC (National Astronomical Observatories of China) Telescope (TNT) at Xinglong Observatory, and the Chinese Ground Follow-up Telescope (CGFT) at Jilin Observatory started to work collaboratively with the GWAC-N as external partners by taking advantage of the ToO alert processing and managing capability of AOM. The GWAC-N can functionally perform the observation tasks to meet all the scientific requirements of the network by adopting the AOM as of the date of this writing (December 2020). A complete GWAC-N will comprise 9 mounts equipped with 36 JFoV and 9 FFoV cameras and several associated follow-up telescopes. Two world-wide sites and advanced CMOS detectors are foreseen to be applied to the GWAC-N in a near future.  The development timeline depends on the future funding and maturity of new technology. Since the GWAC-N is still under development and evolution, this paper describes the structure of network based on the current stages. 

In this paper, we present the GWAC-N's telescopes, the AOM system and the opportunities / science outputs from the GWAC-N. The remainder of the paper is organized as follows. Section \ref{s:gwac-n} describes the system structure of the GWAC-N and the instruments of the GWAC-N, including internal telescopes and the extend partners. In Section \ref{s:aom_system} We then present the AOM of the GWAC-N that we developed for performing for the multi-purpose, flexible, highly efficient observations.  We will describe the scientific opportunities of the GWAC-N and achievement of the network in Section \ref{s:scientific_goal}.  
In Section \ref{s:summary}, we summarize the current status of the GWAC-N and describe network prospects for the near future.

\section{System Structure of the GWAC-N}\label{s:gwac-n}

\begin{figure*}
\centering
\includegraphics[width=0.8\textwidth]{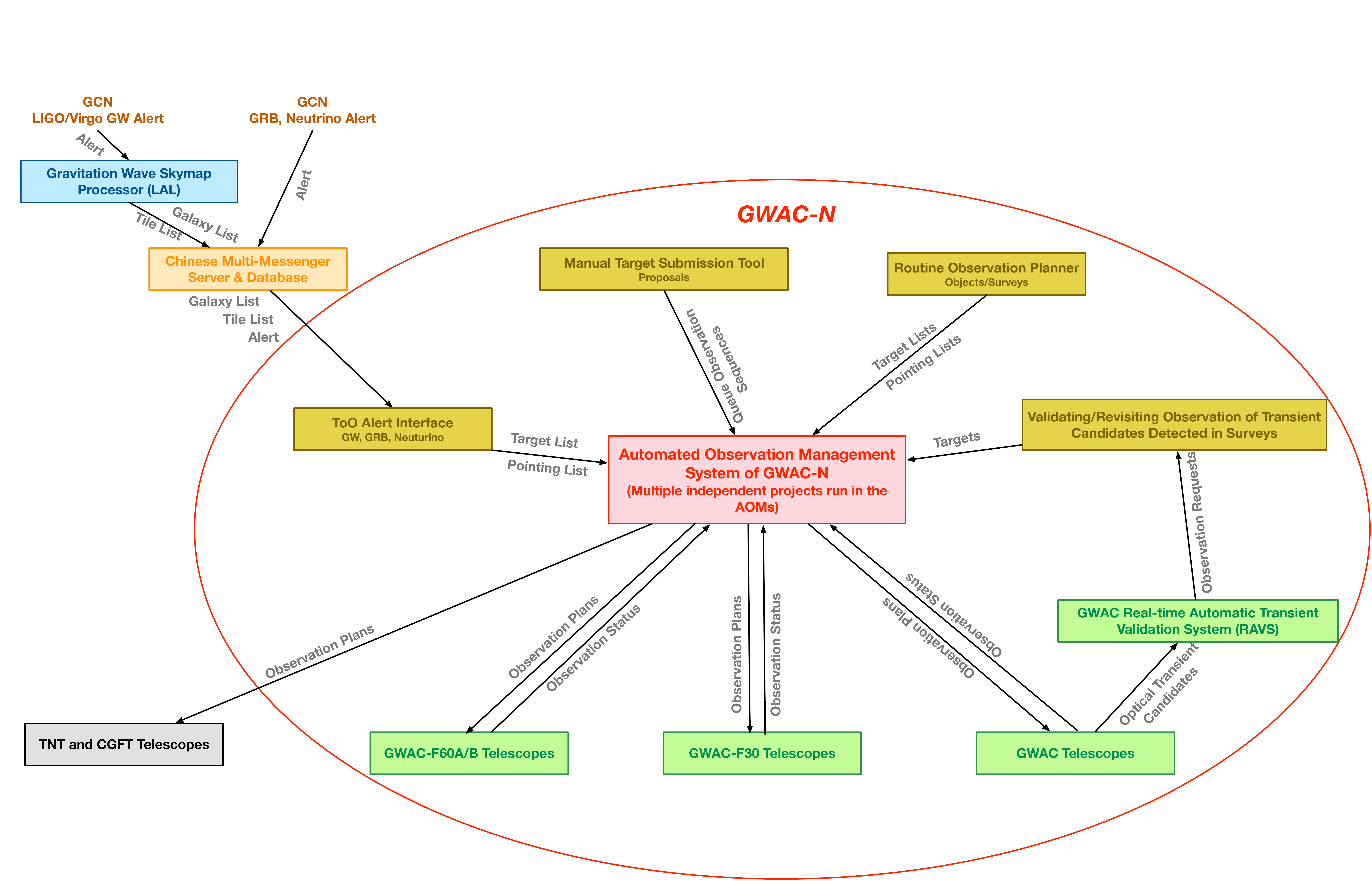}
\caption{ The GWAC-N is composed of interfaces for the target inputing, the multiple telescopes and 
the AOM system. The ToO alert interface can receive the alerts of GW, GRB and neutrinos from the external CMM server . 
The TNT and the CGFT telescopes are connected with the GWAC-N as external partners. 
  }
\label{fig:gwac_architecture}
\end{figure*}

The whole system of the GWAC-N (shown in the Figure 
\ref{fig:gwac_architecture}) comprises three main parts: 
the target input interfaces, the AOM system and the telescopes. In this section, we describe the target 
input interfaces and the telescopes. The AOM system is described in the Section~\ref{s:aom_system}.

\subsection{Target Input Interfaces}\label{ss:target_interface}

The GWAC-N provides multiple external interfaces connected with a variety of alert streams, survey/catalogue planners, the GWAC-A self-detected transient validation system (Xu et al. 2020A) and scientists. All automatic or manual observation requests are inserted into the system via these external interfaces. 

During the O3 run of LIGO/Virgo, the SVOM team develops the Gravitation Wave Skymap Processor (GWSP) at Ir$\grave{e}$ne Joliot-Curie laboratory (IJCLab at CNRS/IN2P3), France.  The GWSP digests the GW skymap and optimizes the tiling observation strategy based on the telescope parameters for the GWAC-A, the GWAC-F30 and the CGFT. Using the Mangrove galaxy catalog (Ducoin et al. 2020A), the GWSP can create optimized galaxy lists for small FoV telescopes like the GWAC-F60A/B. The format of the tiling coordinates and galaxy lists are standard, therefore, it can also be applied to other telescopes, i.e. the GRANDMA (the Global Rapid Advanced Network Devoted to the Multi-messenger Addicts) network (Antier et al. 2020). The tiling and galaxy lists are sent to the Chinese Multi-Messenger (CMM) server using the  VOEvent protocol via brokers. The CMM Service can receive the GRB or Neutrino alert streams from GCN public access by using the pygcn code (Leo Singer \footnote{https://github.com/lpsinger/pygcn/}).
The GWAC-N provides an interface to automatically receive the GW alerts from the CMM in real time. 

Several observation planning codes are running to create target/pointing list for all telescopes to perform routine observations. Each planner can insert the target/pointing list into the AOM using a client provided by the GWAC-N.  The GWAC-N also accepts observation applications from scientists. A tool allows the scientists to customize the observational parameters and to generate complex observation programs. The GWAC-N has another type of targets, the self-detected transient candidates of the GWAC-A validated by the Real-time Automatic transient Validation System of the GWAC-N (RAVS, Xu et al. 2020A). The target needs to be quickly identified and followed-up by the GWAC-F60A/B. Therefore, an interface has been developed for real-time communications between the RAVS and the AOM. 

\subsection{The telescopes}\label{ss:telescope}

The GWAC-A telescopes are the main instruments of the GWAC-N.  Two GWAC-A telescopes are being operated (two more are under testings) at the Xinglong Observatory (lat = 40$^{\circ}$23'39"N, lon = 117$^{\circ}$34'30"E) and founded by the National Astronomical Observatories (NAOC, Chinese Academy of Sciences).
Each GWAC-A mount is equipped with two types of cameras:

$\bullet$ the Joint Field of View (JFoV) camera, a refractive lens with an aperture of 180 mm, is equipped with 4k x 4k CCD camera. The FoV of a JFoV camera is $\sim$ 12.8$^{\circ}$ x 12.8$^{\circ}$. The CCD camera is composed with a 4K E2V chip and a customized liquid cooler system, which allows the CCD works in -50$^{\circ}$ Celsius with respect to the local environment temperature. 4 cameras are installed on a connection frame with angle adjustment mechanism. By carefully adjusting the pointing angles of the JFoV, the four JFoVs cameras are paved in a square sky field. The joint field of view for one mount (four JFoV cameras) reaches about 25$^{\circ}$ x 25$^{\circ}$. The limiting magnitude of the JFoV camera reaches R $\sim$ 16 magnitude for a single image (10 seconds of exposure) in a dark night without cloud. From the stacking images, a typical limiting magnitude of R $\sim$ 18 magnitude is obtained. 

$\bullet$ the Full Field of View (FFoV) camera, a SIGMA 50mm F1.4 lens with aperture of 3.5 cm, is equipped with an Apogee U9000X 3k x 3k CCD camera. The FoV of a FFoV camera is $\sim$ 30$^{\circ}$ x 30$^{\circ}$, which covers the approximately same sky field of the joint FoV of the four JFoV cameras.The FFoV carries out guiding and extending the optical flux coverage to R $\sim$ 6 magnitude at bright end. 

\begin{figure*}
\centering
\includegraphics[width=0.49\textwidth]{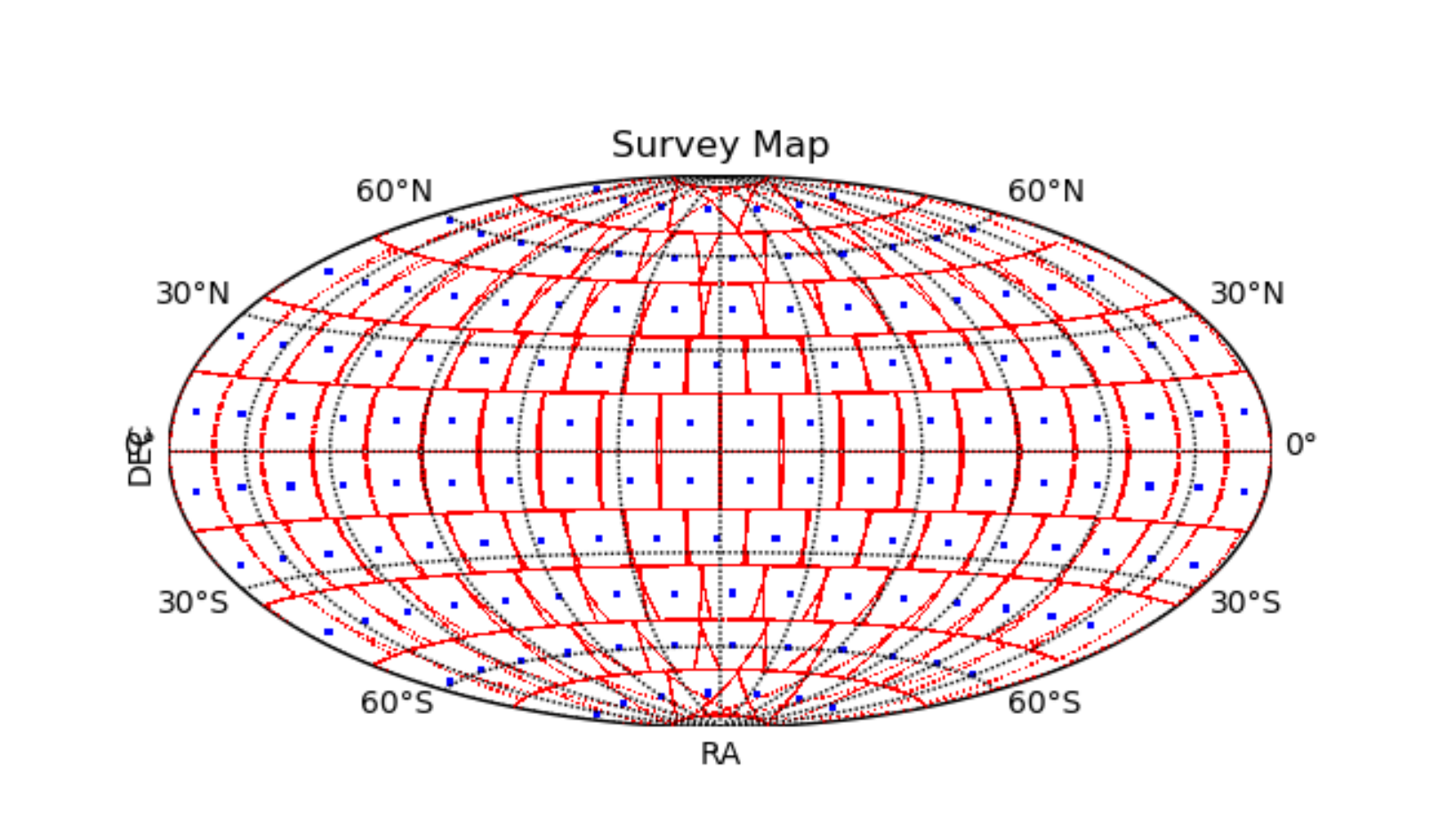}
\caption{ The sky, in Equatorial coordinates, is partitioned into 148 grids of equal area fitting the mount's FoV.  The GWAC-A telescope points to the center of a grid (blue dot), so it can cover the sky field (red square) with its wide FoV. }
\label{fig:gwac_grid}
\end{figure*}

Both types of cameras work on the clear band and variable image cadences depending on objectives (15 or 25 seconds typically). An automatic focusing mechanism developed by Huang et al. (2015) is used on the JFoV and the FFoV cameras to keep the images at their best qualities during the observations. We define a pre-planed grid format, which the all sky is partitioned into 148 fixed grids whose sizes fit the GWAC-A mount's FoV, see Figure \ref{fig:gwac_grid}. The grid format is adopted for the GWAC-A telescopes to carry out all types of observation modes. 

The remote controlling datalink is available for these automatized GWAC-A telescopes. The above features make the GWAC-As being well suited for the optical follow-up of multi-messenger events. The real-time catalog cross-matching and stacking image analysis and transient classification pipeline give to the GWAC-A the capabilities of independently detecting both fast and slow optical transients. 

Two robotic GWAC-F60A/B telescopes and one robotic GWAC-F30 telescope are installed inside of the GWAC dome. The GWAC-F60s are used for the automatic validation of the GWAC OT candidates. They have $\sim$ 10 ${degree/second}$ slewing speed and 18'x18' of FoV with the 2Kx2K Andor iKon-L 936 CCDs. The GWAC-F30 with a FoV of 1.8$^{\circ}$x1.8$^{\circ}$ can complete the gaps of flux coverage and the FoV between the GWAC-A and the GWAC-F60A/B. All the three telescopes are equipped with the Johnson UBVRI filters. With remote controlling and real-time data processing, they can be integrated into the GWAC-N. As a whole, the GWAC-N obtains the capabilities for multiple objectives from survey, queue observations to follow-up observations for many types of targets. The parameters of each type of telescope are summarized in Table \ref{tab:telescope_parameter}.

\begin{deluxetable*}{ccccccc}
 \tabcolsep 0.4mm
 \tablewidth{0pt}
 \tabletypesize{\scriptsize}
 \tablecaption{Telescope parameters \label{tab:telescope_parameter}} 
\tablehead{ \colhead{telescope} & \colhead{number} & \colhead{aperature} & \colhead{FoV} & \colhead{filter} & \colhead{limiting magnitude} & \colhead{number of cameras}  \\
\colhead{} & \colhead{} & \colhead{(cm)} & \colhead{} & \colhead{} & \colhead{(single/stack)} & \colhead{}  } 
\startdata
GWAC & 2 & 18 (JFOV) & 12.5$^{\circ}$*12.5$^{\circ}$ . & Clear & 16/18 & 8 \\
& & 3.5 (FFOV) & 25$^{\circ}$*25$^{\circ}$. & Clear & 12 & 2 \\
\hline
F60A/B & 2 & 60 & 18'*18' & Clear, UBVRI & 18/19 & 2 \\
\hline
F30 & 1 & 30 & 1.8$^{\circ}$*1.8$^{\circ}$. & Clear, UBVRI & 16.5/17 & 1 \\
\enddata
\tablenotetext{}{The limiting magnitude is measured in a 10-second, R band, single image and stacked images. 
The figures of above parameters are up to date as of Dec. 2020.}
\end{deluxetable*}

By using customized datalink, the GWAC-N can collaborate with external telescopes to extend the network. 
The external network currently includes two telescopes: the 80-cm Cassegrain reflecting TNT telescope located at the Xinglong Observatory 
of NAOC and
the 1.2-meter CGFT at the Jilin Observatory of NAOC.  The parameters of the TNT
can be found in Zheng et al. (2008), Huang et al. (2012). The parameters of the CGFT is being tested, 
as the telescope is under hardware updating.


\section{AOM system}\label{s:aom_system}

\begin{figure*}
\centering
\includegraphics[width=0.8\textwidth]{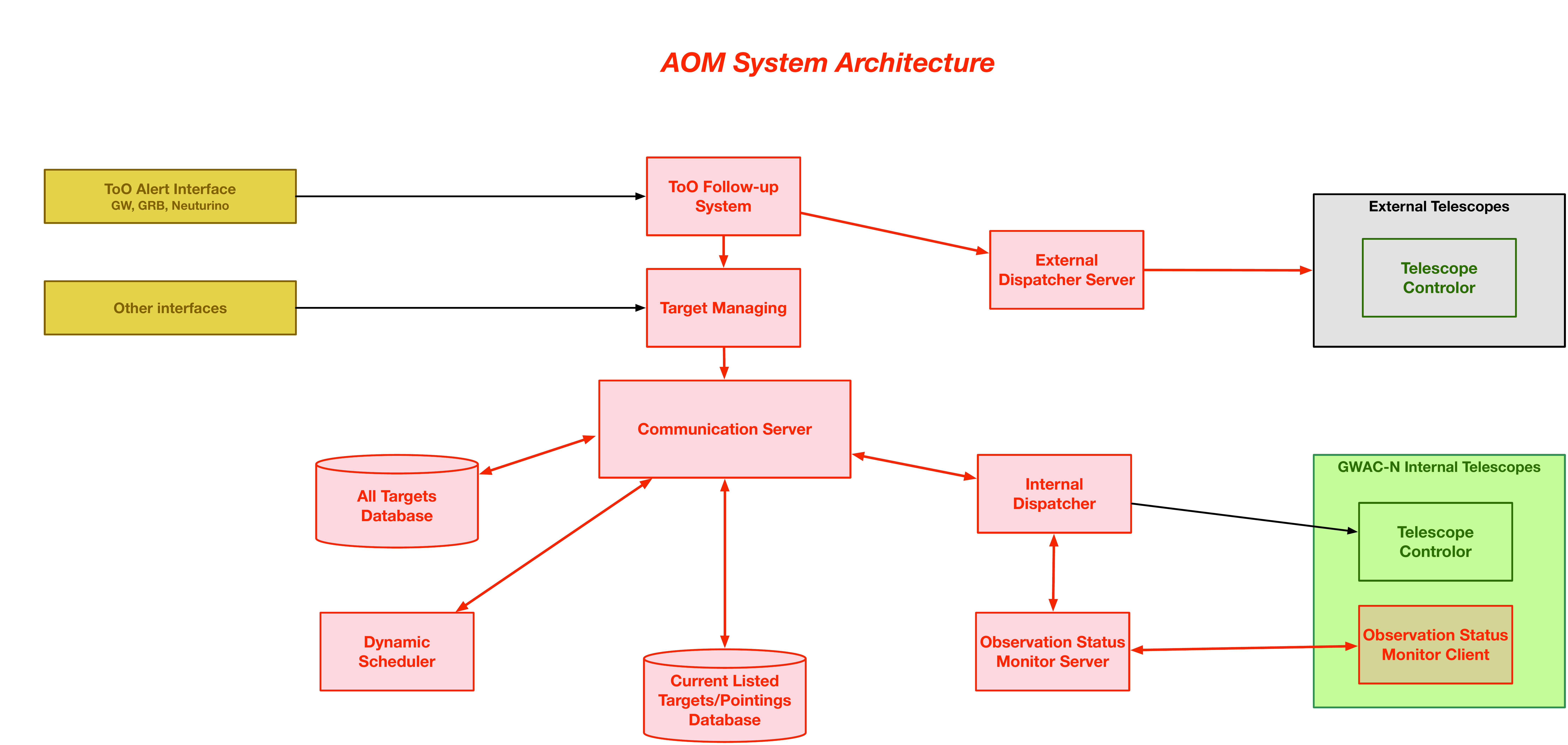}
\caption{  All the sub-systems or modules of the AOM system are shown in the red color. The ToO follow-up sub-system 
obtains the alerts of the ToO from the CMM database as input targets. The target management sub-system receives all other types of 
observation requests, and converts them as input targets. For the internal telescopes of the GWAC-N, a client running in the telescope side
can monitor the observation and data statues, and transmit them back to the AOM. For the external telescopes, the datalink is 
customized for different telescopes. No client is installed in the telescope side in current stage.   
  }
\label{fig:aom_architecture}
\end{figure*}

For a highly efficient telescope network, a strong and smart observation management is a key factor. The automatic observation management is the only way to integrate tens of 
telescopes into a complete network, the GWAC-N. Otherwise, the huge workload during observations is unacceptable for our scientists on duty, not to mention the slow response and
 inefficient observation. Therefore, we developed the AOM system for the GWAC-N to manages all input targets, distributes them to all telescopes and organizes observations with 
all types of strategies.  The system consists of the following sub-systems:  the ToO follow-up, the target management, the scheduler, the dispatcher 
sub-systems and the communication center. The architecture of the AOM is shown in the Figure \ref{fig:aom_architecture}. 
The functions of each sub-system is given in the following sub-sections. 

\subsection{ToO follow-up sub-system}\label{ss:too_followup}

The ToO follow-up sub-system monitors the CMM database for the newly arriving alerts. Currently, 
three types of events, including the LIGO/Virgo GW, the Swift GRB and the Fermi GRB, are selected by the ToO follow-up 
sub-system. 
The sub-system generates a target or a sequence of targets with observation
parameters for the alert that meets the alert selection criteria. 
The observation parameters (such as instrument, observation mode, exposure, etc. ) are set 
based the observation strategies dedicated to different cases.
The alert selection criteria, observation strategies are defined regarding the 
physical variable behavior of the target and the telescope detection capability to increase the chance of detecting the  optical counterpart of the ToO. 
 The details of the selection criteria and observation strategies will be described in another paper 
  (Han et al. in preparation). For the external partners of the GWAC-N, the datalink is customized for a dedicated telescope.
A dispatcher sends a target or target sequence to the TNT and the CGFT 
regarding an alert of the Swift GRB or the LIGO/Virgo GW event. 
By design, no feedback is returned to the AOM from external partners in current stage. A two-side datalink is planned to be
implemented between the GWAC-N and the CGFT in the next stage.

\subsection{Target management sub-system}\label{ss:target_management}

\begin{figure}
\flushleft
\includegraphics[width=.49\textwidth]{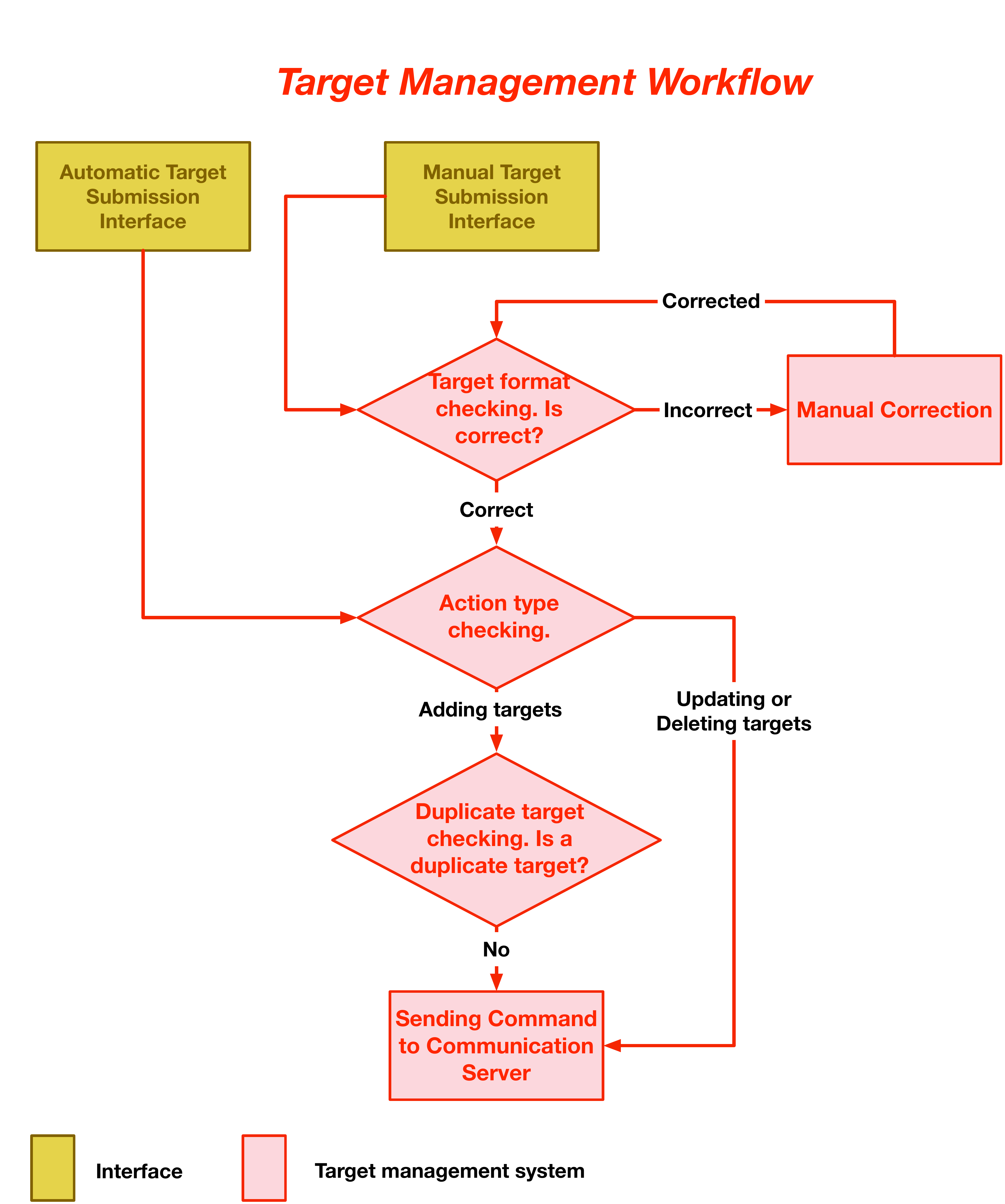}
\caption{ The workflow of target management starts from the target inputting via automatic/manual submission interfaces.
The format validation, manual correction, adding, updating, deleting of targets and duplicate target checking are 
supported by the target management sub-system.
  }
\label{fig:target_management}
\end{figure}

The target management sub-system is to manage the inputs 
from all interfaces to prevent conflicts or duplication processes during the target inputting. The workflow of the target management sub-system is shown in the 
Figure \ref{fig:target_management}. The sub-system automatically checks the format of the inputs and allows scientists and operators to make the corrections. 
A target input message can be either adding new target or updating, deleting target from the sub-system. 
During observation and testing, different interfaces or different users could attempt to repeatedly input targets into the sub-system. 
These duplicated inputs will be rejected by the sub-system to avoid the waste of the telescope resources. 
On the other hand, the sub-system allows the users to perform the repeated observations for a target by adopting a specific observation mode. 

\subsection{Dynamic scheduling sub-system}\label{ss:dynamic_scheduling}

\begin{figure*}
\centering
\includegraphics[width=.9\textwidth]{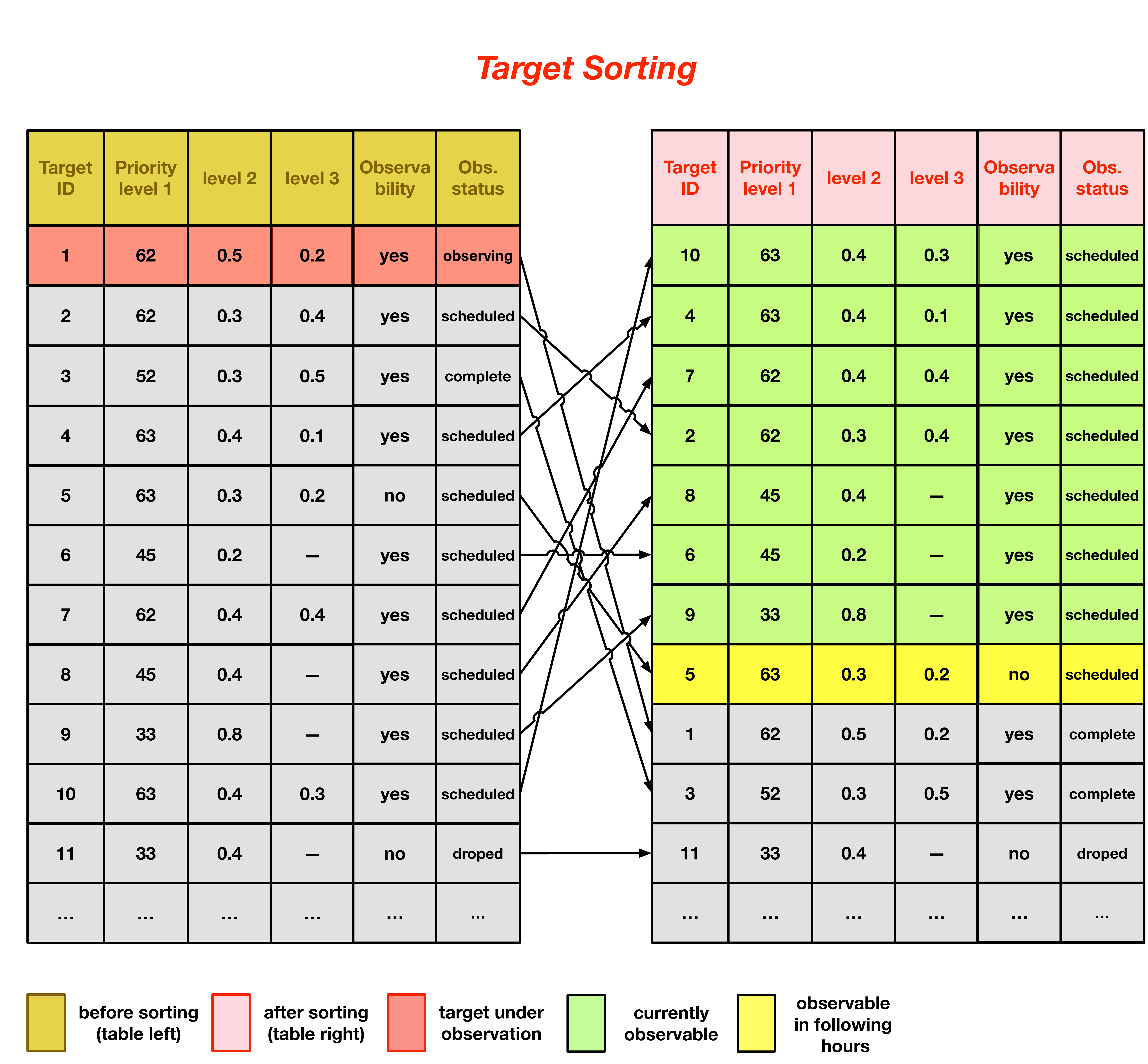}
\caption{  The original target list is shown in the left table. The sorted target list is shown in the right table. 
The first target in the right table enters the observation procedure.
The target under observation is in the red row. The targets observable in the moment of scheduling are 
in the green rows. The targets that are currently not observable, but will be observable in the following hours of 
the night, are in the rows in the light yellow.
  }
\label{fig:dynamic_scheduling}
\end{figure*}

The goal of the scheduler of the GWAC-N is to dynamically generate observation plans for all the telescopes. The principle of scheduling of the GWAC-N is the prioritization of targets. The scheduler satisfies the observation requests for targets with the highest priorities as much as the telescope resources allow to do. The targets with higher priorities can interrupt observations of targets with lower priorities. Multiple levels of grading standards are pre-defined to deal with the complex relation among the target, the observation mode and the telescope. The top level of them is the observation mode, such as, the manual observation, the automatic ToO follow-up, the calibration etc. In this level, each mode of observation is given a range of scores based on its importance. The standard of grading can be changed from telescope to telescope. For example, if a telescope is preferred for a certain observation mode, the score of this mode will be increased for this telescope. Basically, the grading is following the standard definition shown in Table \ref{tab:obs_priority}. We give each mode a range of priority numbers for different cases. For instance, in the automatic ToO follow-up mode, the priority of updated GRB alert is higher than the priority of the initial alert. 
We define the second and/or third level priorities to indicate the sequence of all targets with the same priority in the top level. For different observation modes, the second and third levels can refer to different parameters. Here, we describe two strategies to define the second and third level priorities as examples. For the ToO follow-up observations mode, the rankings need to be adopted in both tiling and galaxy targeting strategies. In our system, the probabilities of tiles and galaxies are defined as the second level priority, while the altitude angle is the third level. For the validation mode, the trigger time (receiving time of target in the system) is the second level priority. No third level is needed in this mode. Using these methods, we can generate observation plans with many complex strategies. 
The scheduler makes a re-sorting process for target list for each telescope based on the priorities, the observability, the status of observation and the telescope, when any update for the target is made in the database. The tables in the Figure \ref{fig:dynamic_scheduling} demonstrates the sorting sequence for the target list during observations. The most important target is listed in the top of the table in the right side. All targets observable in time after the re-sorting are shown in the green cells. The yellow row shows the target being observable later on. Other targets including the ones already completed or no observation time window in that night are not scheduled, which are shown in grey. After re-sorting, the observation plans are refreshed with new order of target list and probable new observation parameters. Each time, the dispatcher picks up the first target in the list for a given telescope. The priority, observability, status of observation and telescope are constantly updated by other sub-systems, which makes the scheduling dynamically.   

\begin{deluxetable*}{l  l  l l l  }
\tabcolsep 0.4mm
\tablewidth{0pt}
\tabletypesize{\scriptsize}
\tablecaption{The grades of observation modes\label{tab:obs_priority}}
\tablehead{\colhead{Observation mode} & \colhead{} & \colhead{Priority} & \colhead{Telescope} & \colhead{Note} }
\startdata
routine mode && 10-19 & GWAC & including surveys with GWAC \\
\hline
normal target mode && 20-29 & GWAC-F60, GWAC-F30 & \vtop{\hbox{\strut including automatic and manual monitoring}\hbox{\strut of targets and supernova survey}} \\
\hline
normal queue mode && 20-29 & GWAC-F60, GWAC-F30  & \vtop{\hbox{\strut including queue observation for periodic objects}\hbox{\strut }}  \\
\hline
automatic validation mode && 40-49 & GWAC-F60, GWAC-F30 & \vtop{\hbox{\strut including automatic validation observations}\hbox{\strut of the self-detected targets of GWAC}} \\
\hline
manual validation mode && 50-59 & GWAC-F60, GWAC-F30 & \vtop{\hbox{\strut including manual validation observations}\hbox{\strut of the self-detected targets of GWAC}} \\
\hline
automatic ToO follow-up mode && 60-69 & GWAC, GWAC-F60, GWAC-F30  & \vtop{\hbox{\strut including automatic follow-up observations}\hbox{\strut for GW, GRB and neutrino}} \\
\hline
manual ToO follow-up mode && 70-79 & GWAC, GWAC-F60, GWAC-F30  & \vtop{\hbox{\strut including manual follow-up observations}\hbox{\strut for GW, GRB and neutrino}} \\
\hline
revisit ToO follow-up mode && 20-29,30-39,80-89 & GWAC-F60  &  \vtop{\hbox{\strut including revisit observation for
interesting targets}\hbox{\strut }} \\ 
\hline
calibration mode && 90-99 & GWAC, GWAC-F60, GWAC-F30  &  \vtop{\hbox{\strut calibration observation for instruments}\hbox{\strut }} \\ 
\hline
manual mode  && 100+ & GWAC, GWAC-F60, GWAC-F30  & \vtop{\hbox{\strut including manual controled observations}\hbox{\strut for all telescopes}} \\
\enddata
\tablenotetext{}{}
\end{deluxetable*}

\subsection{Dispatching  sub-system}\label{ss:dispatcher}

The scheduler does not send any observation command to the telescope controller. A dispatching system does that. Two types of dispatchers are developed for the telescopes inside of the GWAC-N and for the external partners. The technologies used for the external dispatchers depend on the interfaces of the external telescopes. 

\begin{figure*}
\centering
\includegraphics[width=0.98\textwidth]{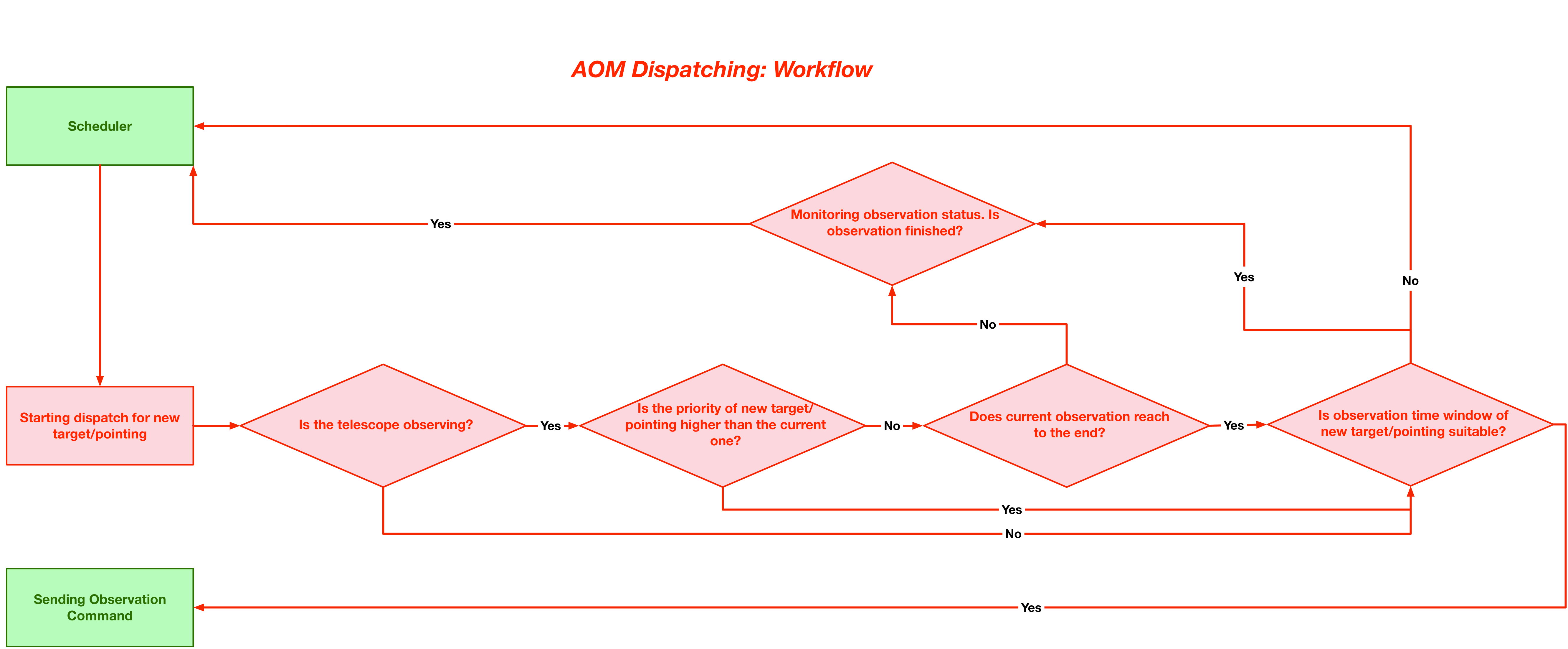}\\
\includegraphics[width=0.98\textwidth]{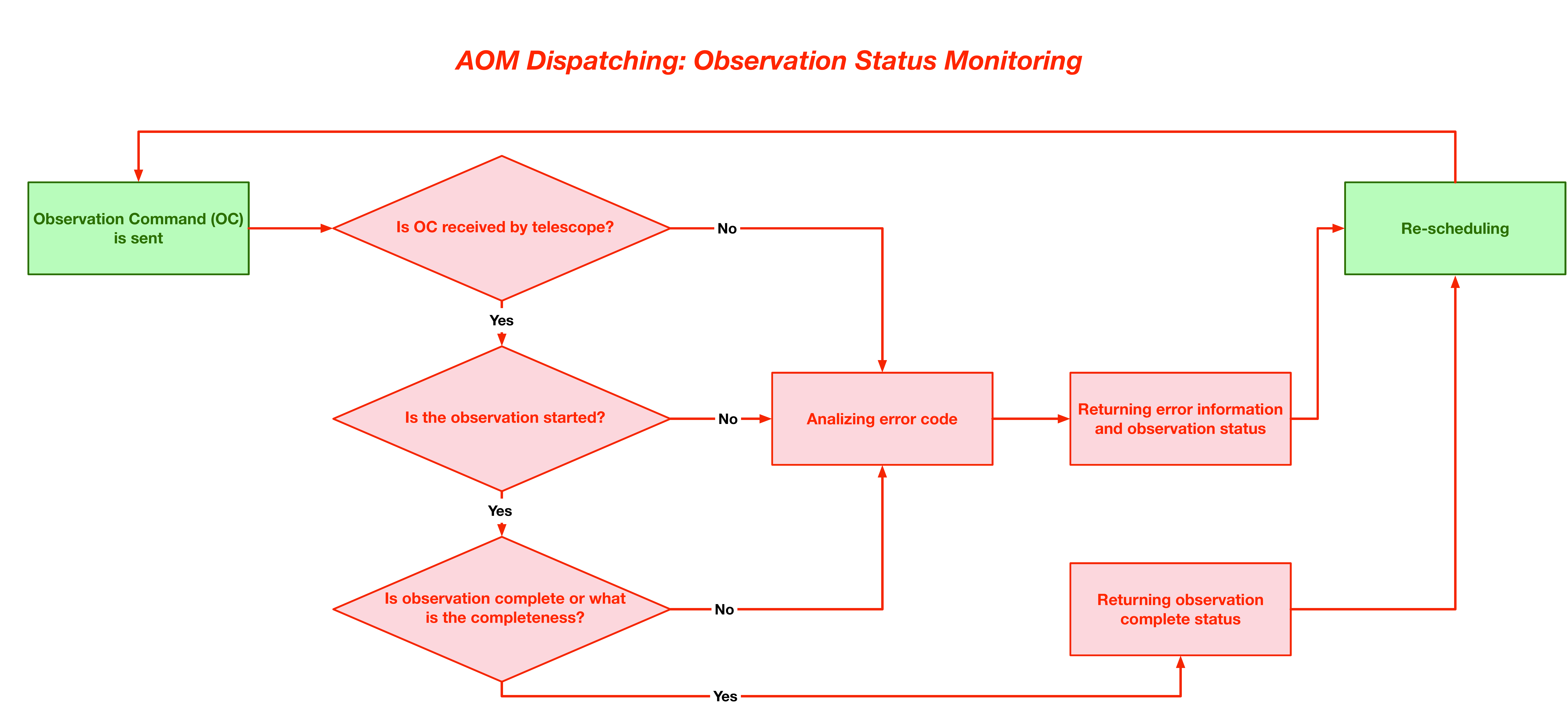}
\caption{Top: The workflow of the AOM dispatching procedure. The tasks of dispatcher are drawn in red color, while scheduler and observation controller sub-system are marked in green. 
Bottom: The workflow of the AOM observation status monitoring procedure, which is drawn in red. 
  }
\label{Fig:dispatching}
\end{figure*}

The internal dispatcher is dealing with the telescopes inside of the GWAC-N. The observation commands are sent to the telescope controller via a one-way link. The observation status is obtained through a link between the monitor server and a client. The dispatcher starts multiple threadings to different telescopes. The work flow is shown in the Figure \ref{Fig:dispatching}. The dispatcher gets an observation plan from scheduler and then it will check the availability of the assigned telescope. If the telescope is available, the dispatcher will check the observability of the target. If yes, an observation command will be sent to the telescope controller. For another case, when the telescope is under observation, the dispatcher compares the priorities of targets. The new target with higher priority can interrupt the on-going observation of an old target. The dispatcher constantly monitors the observation status feedbacked from the observation status monitor. The actions of the dispatcher are based on that real time status. 

The client of the observation status monitor running on the telescope side sends the status back to the server, including the command reception, observation status and completeness. An error code is also sent back to the server, which can be used for system error analysis.  

\subsection{Communication center}\label{ss:communication_center}

\begin{figure*}
\centering
\includegraphics[width=.98\textwidth]{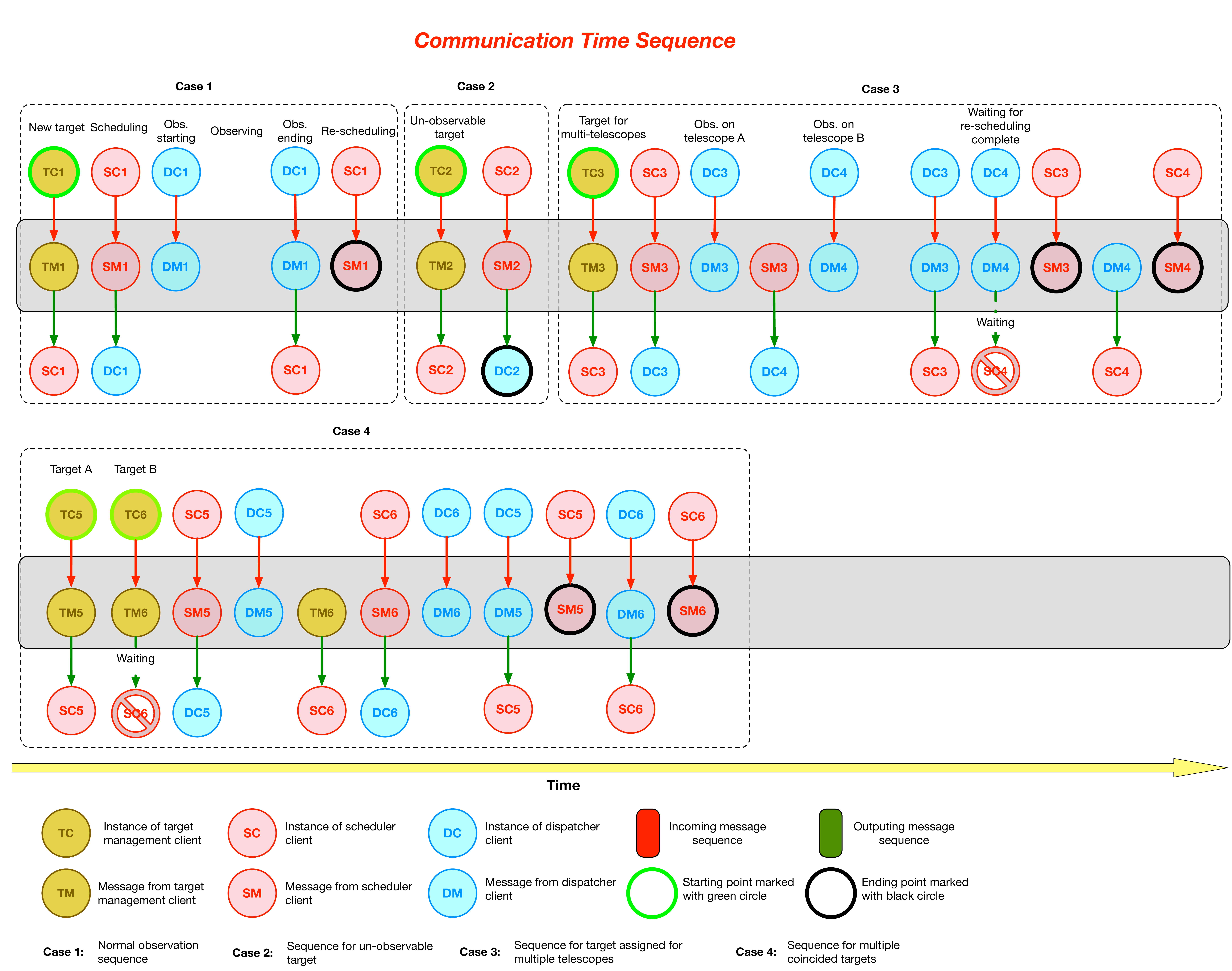}
\caption{ The figure demonstrates the four typical scenarios of the message sequences in the AOM system. A functioning communication center (CC) consists of a message sequential controller, multiple instances that one of each is launched and communicates with a given connected client. The message sequential controlling of the CC is shown in the grey box. The incoming and outputting message flows for the controller are drawn in red and green colors respectively.  To clarify the message sequencing procedure, we use yellow, red and blue colors to mark the actions of target management (shorten with TC and TM), scheduling (SC and SM) and dispatching (DC and DM).  A green circle and a black circle represent the starting point and the ending point of a message flow.  
  }
\label{Fig:communication_time_sequence}
\end{figure*}

The AOM system is composed of many sub-systems and a database. The communications are very complex and frequent among external interfaces, sub-systems and telescopes. To avoid the conflicts and sequential confusion in the communications, a sequential control mechanism is crucial for the AOM system. In the earlier version, all sub-systems are directly communicating to the database. When a large number of concurrency entries occur in the database, a protecting mechanism will be triggered in the databases from preventing damages of data. These chance faults are rare but fatal to our system. Another key point is for the scheduling. Unlike the multi-instance dispatcher, only one instance of scheduler can be run at time, because to deal with the dynamic information by multiple schedulers easily causes the information confusion. The AOM system must ensure that the scheduling is well organized in such a complex situation. A sequential controller can solve those communication issues. Therefore, we developed a communication center (CC) combined with communicating and sequential controlling functions and a communication client deployed on each sub-system. The CC runs a server and many instances (communication modules). An instance is launched when a connect request is created by a server or a client running on a sub-system. All messages between the server and the clients are marked with flags to indicate different types of the messages. In the server side, the messages will be classified and distributed to the dedicated clients in the proper orders. The procedures of observation scheduling and dispatching depend on the ordering of the messages. In the client side, each message will be treated as an independent message, and be processed only in order of arrival. In this paper, we simulate four scenarios to show how the observational procedures are executed smoothly in the GWAC-N. These scenarios are the most typical cases of communication time sequences during the scheduling and dispatching (see the Figure \ref{Fig:communication_time_sequence}):

\smallskip \textbf{Case 1}. Communication time sequence for normal observation procedure. In the normal case, the procedure starts at the point, which a target is added into the target list. The next steps are the scheduling, the dispatching, and the observation. The final step is a re-scheduling process. To be specific, after a new target is put into the target list from an interface (TC1), the target management client will process it, format it and send a message (TM1) with observation parameters of the target to the CC. The message will be added to a message list organized by a sequencer in the CC. The target message is instantaneously sent to an instance of the scheduler (SC1) that will start to make an observation plan. The scheduler generates the observation plans not only for that new added target but also for all observable targets in the target list. After the scheduling is done, a message with a status of scheduling (SM1) is returned back to the CC, then a command of dispatching will start an instance of dispatcher client (DC1). The dispatcher client decides to choose a target with top priority from the target list for the next observation or to wait the completeness of the current observation. There are multiple instances of dispatcher clients running simultaneously to control different telescopes. The client of the dispatcher sends messages (DM1) to inform the CC when the observation is started and finished. After the observation is done, the scheduler client receives a command from the CC to start re-scheduling to update the observation plans. The instance of dispatcher client is closed then. The procedure ends at this point.   

\smallskip \textbf{Case 2}. When a target is added into the target list, the scheduler will firstly compute the observational time window. The one without the observational time window from TC2 will not be scheduled. The instance of the scheduler (SC2) will still communicate to the CC for the scheduling status (SM2) to inform the dispatcher (DC2) that the update of target list. The procedure ends at the dispatcher (DC2). 

\smallskip \textbf{Case 3}. Multiple telescopes are needed to observe one target. This situation usually occurs when synchronized multi-band photometry is performed for the target. In the Figure \ref{Fig:communication_time_sequence}, we assume that two telescopes are used in that scenario. After receiving the target information from an interface (TC3), the instance of the scheduler (SC3) generates two observation plans for two telescopes respectively. Then the CC starts the first instance of a dispatcher client (DC3), while the second instance of a dispatcher client (DC4) will not be started, until the CC gets the feedback message (DM3, the starting status of observation) from the DC3.  Then the DC4 sends an observation command to the second telescope and observation status message (DM4, the starting status of observation) to the CC. When the DC3 obtains the complete status of the observation, the SC3 will start re-scheduling process. In the meantime, the DC4 obtains the status of the second observation, but the message transmission (from DM4 to an instance of SC4) will be put on hold until the observation plans are refreshed by the SC3. Then the SC4 is started. The procedure is finished when the SM4 is received.

\smallskip \textbf{Case 4}. In this case, dozens or even hundreds of targets/pointings are added into the system at nearly same time. This situation happens frequently during the Multi-Messenger follow-up observations.  We simulate the scenario when two targets (TC5 and TC6) are inserted in the same time.  An instance of scheduler (SC5) is started immediately when the message of target (TM5) is transmitted. The SC5 and a dispatcher client (DC5) are executed successively. The message (TM6 ) for the second target (TC6) will be transferred after the status message (DM5, the starting status of observation for the TC5) is received. Then the second instance of scheduler (SC6) and  a dispatcher client (DC6) are executed for the TC6. When the observations and re-scheduling are finished for the TC5 and the TC6, the procedures end.        

\smallskip In the procedure of above cases, both actions of scheduling and actions of dispatching are triggered by the dedicated messages. The sequential controller can organize the messages in proper orders, which prevents the confliction and sequential confusion during observations.  

\subsection{AOM system workflow}\label{ss:aom_workflow}

\begin{figure*}
\centering
\includegraphics[width=.9\textwidth]{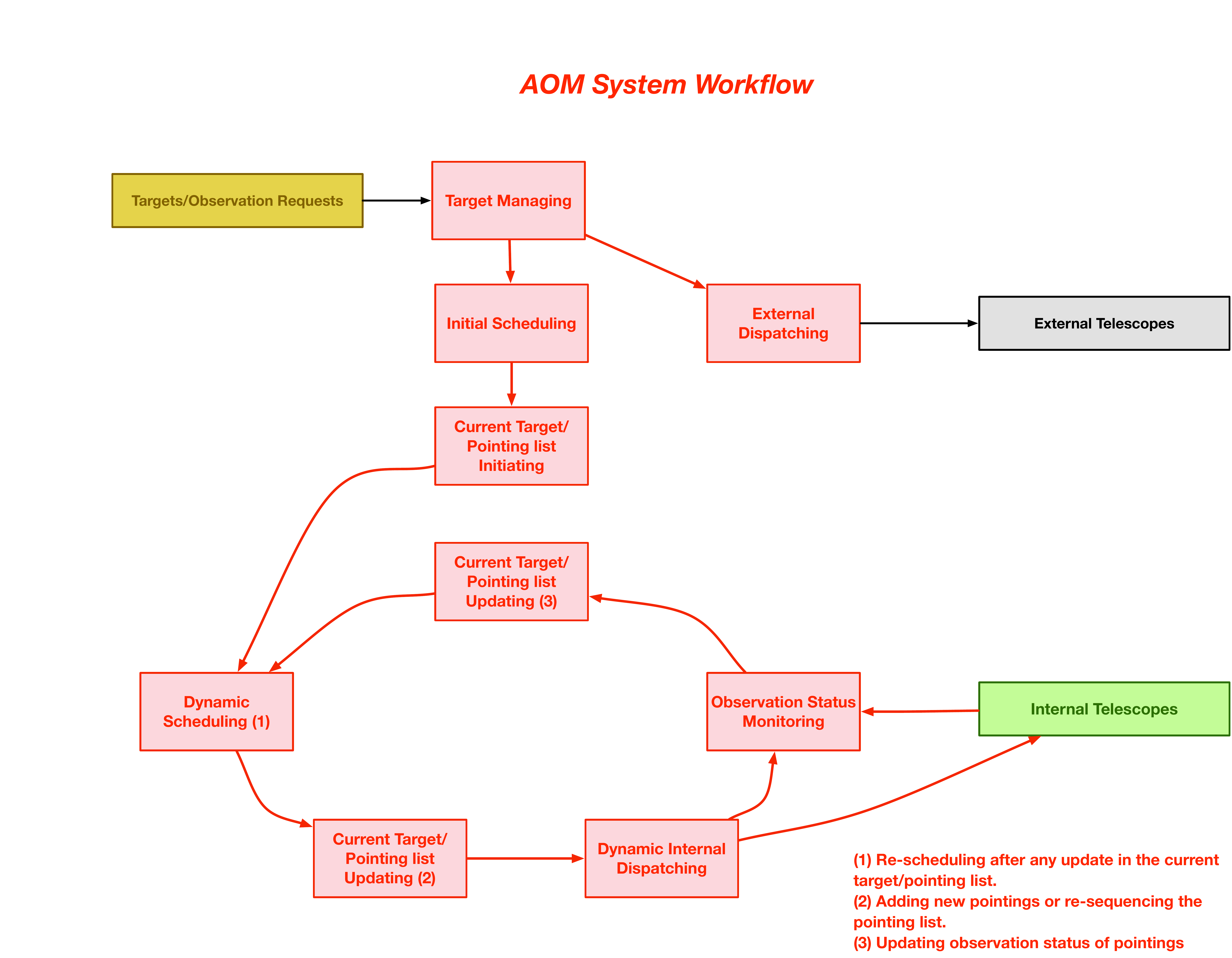}
\caption{ The workflow of a complete observation management sequence controlled by the AOM system that is drawn in red color. The target management sub-system receives all the targets from the interfaces (in yellow). The external dispatcher deals with the external telescopes (in the grey box) of the GWAC-N. The system fully controls the internal telescopes (in the green box) with the scheduling, dispatching and status monitoring sub-systems.  
  }
\label{Fig:aom_workflow}
\end{figure*}

Combining with above sub-systems, the overall workflow of the AOM system is described as follows. The targets are manually/automatically inserted into the system by using interfaces provided by the AOM system (all observation request are treated as targets). All the targets are processed and classified by the target management sub-system, then are inserted into the target database. Some targets are sent by the external dispatcher to trigger the follow-up observations with external telescopes. The targets for the GWAC-N will be initially scheduled in order to compute the observation time windows and to make the initial observation plans. The targets having the observation time windows are added in the daily target list and are stored in the database. This daily target list contains all the targets to be observed in a given night. This list is kept updated during the night, since new targets come in, the observation parameters and status of targets are updated, and some targets are removed from the list. Triggered by the CC, the dynamic scheduler makes observation plans for the target in the target list, and the dispatcher selects a target to observation command based on the observation plan, status of observation and the status of telescopes. The observation status monitor keeps observation status updated in the target list, so the scheduling and dispatching can be fully dynamically. The system completes a closed control loop (shown in the Figure \ref{Fig:aom_workflow})

\subsection{Performance of the AOM system}\label{ss:aom_performance}

With the AOM, the GWAC-N integrates 5 telescopes and collaborates with 2 external telescopes. By using the standard link provide by the AOM, the telescopes can easily join the network. The AOM also provides the customized link for external telescopes, which is very useful for those telescopes willing to join the ToO follow-up campaigns without developing their own follow-up system.  The AOM can perform complex observations with currently 10 observation modes and 175 strategies. To add or modify the observation modes or strategies, the users only need to edit the configuration file rather than change the code of the system. During the operations, the routine survey, the target monitoring, the GWAC OT validation and the ToO follow-up observations are done automatically. The operators are needed for monitoring the operation status and manually importing the targets with the special observation requirements into the system.  

The communication mechanism and system structure of the AOM ensure the stability of the system, which is another key factor for a robotic telescope network. The AOM has been implemented in the GWAC-N since August 2019. During the observation season of 2019-2020 (from October to April, weathers are better and night time is longer comparing with the rest of year), the AOM is working with a high duty cycle and a stable behavior. The AOM produced 622 observation plans per clear night on average (in December 2020), with failure-free. On the 7th of December, the AOM produced 1064 observation plans, which is the highest working load in the month. 

The efficiency of the AOM can be valued by the time delay between the target inserting and the observation command sending. Shortening the time delay is very important for a ToO orientated observation network. The time delay is usually caused by the communication delays, the scheduling process and the observation status monitoring process. In the AOM, the main time consuming is from the scheduling process, which depends mainly on the number of targets in the target list. Because once an observation is finished, the AOM will start a re-scheduling process for all the targets in the list and send an observation command to a telescope. During December 2020, the list contains an average number of 1500 targets in a night. On the 7th of December 2020, the AOM handled 3955 targets, which is the largest number for a night among the month. We tested the time delay of each re-scheduling process during the night. The longest time delay is less than 2 seconds, which is negligible comparing with the time delay in the telescope side between stopping the previous exposure and starting new observation. 
We simulated the extreme scenario of observing 10000 targets for 10 telescopes. The time delay is less than 4 seconds, which makes the AOM qualified for all types of the ToO follow-up tasks and for a large telescope network. 


\section{Scientific Opportunities and Output to the GWAC-N}\label{s:scientific_goal}

The primary goal of the GWAC-N is to observe the prompt emission of GRBs in optical bands. We emphasized the role of the telescopes of the GWAC-A because of their key features of large sky coverage, high time resolution and real time transient detection capability. These features allow the GWAC-A to independently search for optical transients with a high cadence. Furthermore, the GWAC-A can be also used for follow-up observation of multi-messenger event. The associated multi-band small FoV telescopes in the GWAC-N are originally designed for the real-time automatic validation for the optical transients detected by the GWAC-A.  These telescopes can be also used for other purposes, such as photometry of variable object, galaxy targeting observation for multi-messenger events and supernova survey. 

\subsection{Gamma-Ray Burst}\label{ss:GRB}

The prompt emission of GRB in optical bands is difficult to be observed, since its very fast temporal decay. To obtain the prompt emission, the speed of  response of telescopes to a GRB alert is highly desired. The idea behind the design of the GWAC-A is to eliminate the response time to a GRB alert. The total sky coverage of full GWAC-A is as large as 5000 square degrees, which can cover the same sky area being monitored by the ECLAIRs telescope, the main GRB detector of SVOM (Wei et al. 2016). This extreme large sky coverage guarantees that the GWAC-A simultaneously discovers the optical counterparts for about 30\% SVOM detected GRBs at tigger time (T0). It can also make the GWAC-A to be a suitable instrument to follow up the GRBs detected by other gamma-ray instruments, which cannot provide accurate localizations (the Fermi Gamma-ray Space Telescope and SVOM/GRM, etc.). 

The two GWAC-F60A/B telescopes and the GWAC-F30 telescope in the GWAC-N robotically follow-up the GRBs detected not only by SVOM but also by the Swift satellite. Since 2016, these 3 telescopes manually followed up 6 Swift GRBs (Xin et al. 2016, Xin et al. 2017A, Han et al. 2018A, Xin et al. 2019A, Xin et al. 2019B, Xin et al. 2019C). Since 2020, the AOM automatically followed up 3 Swift GRBs by using GWAC-A, GWAC-F60 and TNT telescopes (Xin et al. 2020B, Xin et al. 2020C, Xin et al. 2020D, Xin et al. 2021) . For GRB 201223A, the optical counterpart was detected in a GWAC-A image taken at 2 seconds after the burst. The GWAC-F60A started the follow-up observations for the counterpart 23 seconds after receiving the alert of the burst and 44 seconds after the burst trigger. These observations can provide consecutive lightcurve from the prompt emission phase to the afterglow phase (Xin et al. 2020C, Xin et al. 2020D).   

\subsection{Multi-Messenger Target of Opportunities astronomy (gravitational wave, neutrino)}\label{ss:too-mm}

The poor localization of the Multi-Messenger Target of Opportunities (ToO-MM) alert is a great challenge for all the optical follow-up facilities. To quickly search for the optical counterparts in a large sky area, two observation strategies are widely used by most of the optical telescopes for the ToO-MM follow-ups, which are either by tiling the large localization regions or by performing galaxy-targeted observations. By using all the telescopes by the AOM, the GWAC-N can conduct efficient follow-up observations with both strategies.  Taking advantage of the wide field of view of telescopes, GWAC-A can cover a significant portion of the ToO-MM localization regions in a very short amount of time by using the tiling strategy. In the meanwhile, the GWAC-F60A/B and GWAC-F30 carry out galaxy targeting observations. As a group, three telescopes can search $\sim$500 galaxies in a clear night. During the O2 and O3 GW run, the pathfinder telescopes mini-GWAC array and the GWAC-A performed follow-ups of large sky covering for 25 of GW events (8 in O2 and 17 in O3, Dornic et al. 2019, Ducoin et al. 2020B, Ducoin et al. 2020C, Gotz et al. 2019, Han et al. 2019, Lachaud et al. 2019, Leroy et al. 2017, Mao et al. 2020, Turpin et al. 2019A, Turpin et al. 2019B, Turpin et al. 2019C, Turpin et al. 2019D, Turpin et al. 2020A, Wang et al. 2019, Wang et al. 2020A, Wang et al. 2020B, Wei et al. 2017A, Wei et al. 2017B, Wei et al. 2017C, Wei et al. 2017D, Wei et al. 2017E, Wei et al. 2017F, Wei et al. 2019A, Wei et al. 2019B, Wei et al. 2019C, Wei et al. 2019D, Wei et al. 2020, Wu et al. 2019, Xin et al. 2017B, Xin et al. 2019D, Xin et al. 2019E, Xin et al. 2020A).

\subsection{Optical transient target (supernova, flare)}\label{ss:ot}

Thanks to the large sky coverage of the GWAC-A, the fast follow-up capability of the GWAC-F60A/B and GWAC-F30 and the dedicated online data processing pipeline of each telescope, the GWAC-N is not only capable to independently detect optical transients in the sky but also to identify the types of the candidates in realtime. Since 2018, the GWAC-N has detected several super stellar flares (Han et al. 2018B, Wang et al. 2020C, Xin et al. 2020E).  The GWAC 181229A, a supper flare with an amplitude of $\Delta R \sim9.5$ mag was detected by the GWAC-A and was classified as a ultracool M9 type star by the photometry follow-ups of GWAC-F60A and spectroscopic observations of the NAOC 2.16m telescope (Xin et al. 2020E). This is a good example to demonstrate the capability of the GWAC-N. 

The routine survey mode of the current GWAC-A covers $\sim$20000 square degree of observable sky (galactic plane not included) on a clear night. The center of the sky coverage of the survey shifts $\sim$ 1 degree along the longitude each day, while most of survey area is consistent in successive observation nights, which means a supernova survey with 1 day cadence can be made using the GWAC-A data. We adopt the ASAS-SN supernova detection relation (Fig 6. in Holoien et al., 2017) to estimate the detection rate of the GWAC-A. With a limiting magnitudes of m$_{R}$ $\sim$ 16 in a single image and $\sim$ 200 clear nights per year at the GWAC site, we estimate that the GWAC-A is capable to detect about 30 bright, nearby supernovae per year by using a dedicated pipeline. 

\subsection{Variable and Periodic object}\label{ss:variable}

The most of sky coverages of the GWAC-A's survey are consistent in successive observation nights, which means one given sky area can be monitored for days or dozens of days. With a high cadence observation mode (15 seconds per image), the GWAC-A can monitor the variables or the periodic objects in the sky area and obtain their variation. The online data processing pipeline of the GWAC-A can measure the photometric features for all sources in the images. Using neural network mechanism, researchers analyze the massive data of the GWAC-A to detect and to classify variable and periodic sources (Qiu et al. 2018, Turpin et al. 2020B).  

\subsection{Moving object}\label{ss:moving_obj}

\begin{figure*}
\centering
\includegraphics[width=.9\textwidth]{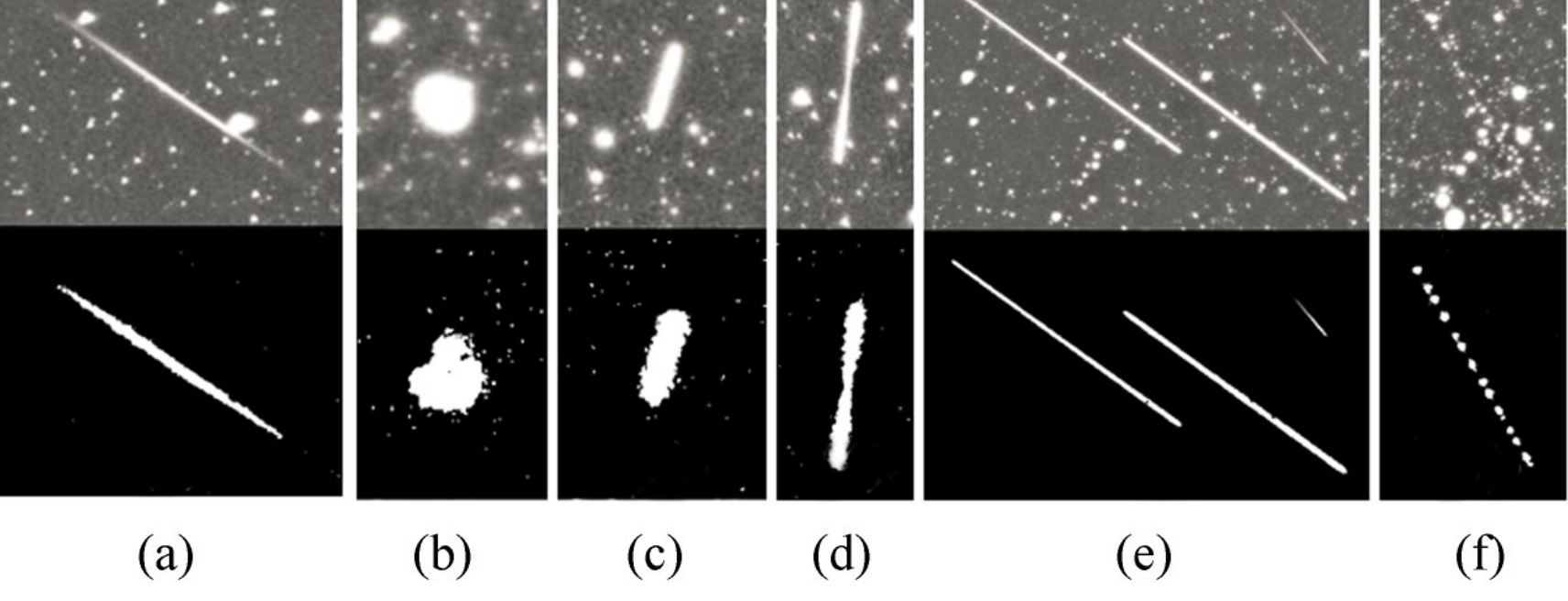}
\caption{Some moving object images of the GWAC-A. The top part of images are original images. The bottom part of images are residual images. (a) is meteor candidate, (b-f) are other types of moving objects including giant planets, minor planets and artificial satellites.
  }
\label{Fig:moving_object}
\end{figure*}

With its large field of view and high cadence, the GWAC-A can monitor hundreds asteroids on an observation night. The GWAC-A also has the capability to detect the decameter asteroids and meteors (Shugarov 2019, Xu et al. 2020B). They are valuable for the researchers in these fields. Our team works on the algorithm and database to recognize and morphology analyze them from the GWAC data. The Figure \ref{Fig:moving_object} shows some moving objects automatically detected in the GWAC-A images by an algorithm for selecting moving objects. The accuracy over 85$\%$ can be reached for the meteor candidate selections by using the algorithm (Xu et al. 2020B). 

\section{Summary and perspective}\label{s:summary}

The GWAC-N is currently composed of two GWAC telescopes, two GWAC-F60 telescopes and one GWAC-F30 telescope. It is also collaborating with two external telescopes: the CGFT and the TNT telescope. By implementing the AOM, those telescopes can work as a network smoothly. Besides of routine observations, the GWAC-N performed follow-up observations of the LIGO/Virgo GW, the Fermi GRB, the Swift GRB events. During the LIGO/Virgo O3 campaign, using the AOM system described in this paper, the GWAC-N observed 17 GW events and published 23 GCN circulars.

In the next two years, the complete GWAC-N will be installed at two observatories. The number of telescopes in the GWAC-N will be extended to nine GWAC-A telescopes and five 60cm class telescopes. More external telescopes are also foreseen to join the network. The AOM will fully support the operations of the GWAC-N and greatly enlarge the scientific return of the GWAC-N.  Its technology and mechanism, or as a whole, the AOM can be adapted to other world-wide, general purposed, telescope networks. The code of the whole AOM system is available at the Github 
\footnote{https://github.com/yunws/GWAC\_obs\_manage\_system/}  
for public download.

The AOM is not only used to manage the operations of telescopes but also to manage the data. A server named the Data Center (DC) is installed inside of the AOM.  In the past, the massive data taken from the GWAC-N telescopes bring huge workload to the scientists and operators to collect and find images for certain observations. Currently, the data of the ToO follow-up observations taken by the GWAC-F60, the GWAC-F30 are automatically collected and uploaded to the DC by the AOM. It allows we centralized the data processing in the DC rather than the data processing distributed in the telescope side.  In the future, we plan to integrate the data processing pipeline in the AOM system, as well as the data product release.

\section{Acknowledgement}

The GWAC team at the NAOC is grateful for financial assistance from the National K\&D Program of China (grant No. 2020YFE0202100) and the National Natural Science Foundation of China (Grant No.
11533003, 11973055, U1831207, 11863007). This work is supported by the Strategic Pioneer Program on Space Science, Chinese Academy of Sciences, grant Nos. XDA15052600 \& XDA15016500 and by the Strategic Priority Research Program of the Chinese Academy of Sciences, Grant No.XDB23040000.  Damien Turpin acknowledges the financial support of the CNES post-doctoral program. We thank the staffs of the Xinglong and the Jilin observatories at which the TNT and the CGFT telescopes are operated. We would like to thank Sarah Antier, David Corre, Jean-Gr\'egoire Ducoin for their very helpful discussions during the developments of the AOM system.




\end{document}